\documentclass[prd,aps,showpacs,preprintnumbers,eqsecnum,amsmath,amssymb,nofootinbib]{revtex4}
\usepackage{bm} 
\usepackage{epsfig}

\begin{document}
\title{Rare radiative leptonic decays ${\bm B_{d,s}\to \ell^+\ell^-\gamma}$} 
\author{
Dmitri Melikhov and  
Nikolai Nikitin
} 
\affiliation{
Institute of Nuclear Physics, Moscow State University, 119992, Moscow, Russia}
\date{\today}
\begin{abstract}
Long-distance QCD effects in 
$B_{d,s}\to\ell^+\ell^-\gamma$ decays are analyzed. We show that the contribution of the light 
vector-meson resonances related to the virtual photon emission from valence quarks 
of the $B$-meson, which was not considered in previous analyses, strongly 
influences the dilepton differential distrubution. 
Taking into account photon emission from the $b$-quark loop, weak annihilation, 
and Bremsstrahlung from leptons in the final state, we give predictions 
for dilepton spectrum in $B_{d, s}\to \ell^+\ell^-\gamma$ decays within the Standard Model. 
\end{abstract}
\maketitle
\section{Introduction}
Rare radiative leptonic $B_{d, s}\to\ell^+\ell^-\gamma$ decays 
are induced by the flavour-changing neutral current transitions $b\to s,d$. 
In the Standard Model such processes are described by penguin and box diagrams 
and have small probabilities  $10^{-8}$ -- $10^{-15}$ (see e.g. \cite{ali}). 
Processes with such small branching ratios cannot be observed at the running 
machines such as Tevatron, BaBar and Belle. The decays $B_{d, s}\to\mu^+\mu^-$ and 
$B_{d, s}\to\mu^+\mu^-\gamma$ will be studied at LHC with the detectors 
ATLAS, CMS, and LHCb \cite{lhc}. These decays are perspective candidates 
for a search for physics beyond the Standard Model, therefore reliable theoretical 
predictions for these decays are of great interest. 

The effective Hamiltonian describing the $b\to q$ ($q=d,s$) weak transition has the form \cite{wils} 
\begin{equation}
\label{heff}
{\cal H}_{\rm eff}^{b\,\to\, q} = \frac{G_F}{\sqrt{2}} V_{tb} V_{tq}^\ast\, 
\sum_i C_i(\mu) \, O_i(\mu),
\end{equation}
$G_F$ being the Fermi constant, $C_i$ the scale-dependent Wilson coefficients, and $O_i$ 
the basis operators. For $B$ decays $\mu\simeq 5$ GeV is a convenient choice. 
The amplitudes of the basis operators between the initial and final states 
may be parametrised in terms of the Lorentz-invariant form factors. For radiative 
leptonic decays several different basis operators contribute, and respectively, one 
encounters several different types of form factors. The latter contain nonperturbative 
QCD contributions, and therefore their calculation poses one of the main problems for 
theoretical analysis of $B\to \ell^+\ell^-\gamma$ decays. 

The $B\to \ell^+\ell^-\gamma$ decays  have been studied in several papers
\cite{mk2003,kitay,sehgal,univ,aliev} where transition form factors have been analysed and 
partial widths, photon energy spectra, dilepton mass spectra, and charge asymetries have been 
calculated. It turns out however that an important contribution related to hadron 
resonance in the $\ell^+\ell^-$ channel was not taken properly into account. Our goal is 
to account for this contribution and to give reliable predictions for dilepton mass spectrum and
decay rates.\footnote{We do not consider any asymmetries since they cannot be studied at LHC 
because of a small expected number of events \cite{lhc}.}   

The paper is organized as follows: In Section \ref{sec:2} we discuss 
the dominant contribution to the decay amplitude related to the transition 
$\langle \gamma|H_{\rm eff}(b\to q \ell^+\ell^-)|B\rangle$. 
In Section \ref{sec:3} we study the contribution related to the 
$B\to \ell^+\ell^-$ transition induced by the photon penguin operator 
$\langle \ell^+\ell^-| \bar d \sigma_{\mu\nu}q|B\rangle$.  For a proper 
description of this process it is necessary to take into account 
light meson resonances which emerge in the physical decay region. 
Making use of the vector meson dominance we relate these contributions 
to the $B\to V\gamma$ transition form factors. 
Weak annihilation is discussed in Section \ref{sec:4}. This process is known to be suppressed 
by $1/m_b$ compared to the photon emission from the $B$-meson loop. 
Nevertheless, it gives sizeable contribution to the dilepton mass spectrum in the region of 
small dilepton momenta. 
In Section \ref{sec:5} we consider Bremsstrahlung from leptons in the final state. 
Its contribution to the cross section is proportional to $(m_{\ell}/M_{B})^2$ and 
is thus essentail mainly for $\tau$ leptons in the final state. 
In Section \ref{sec:6} we give numerical predictions for the branching ratios and dilepton mass 
spectra based on our detailed analysis of the form factors.  

\newpage
\begin{figure}[tbh]
\begin{center}
\epsfig{file=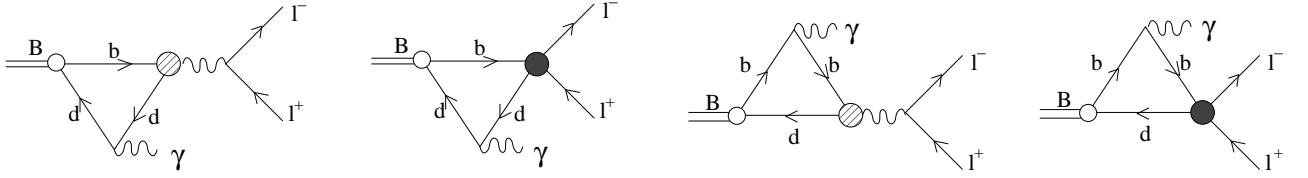,width=17.0cm}
\caption{\label{fig:1} Diagrams contributing to 
$B_d\to\ell^+\ell^-\gamma$ discussed in section \protect\ref{sec:2}.
Dashed circles denote the $b\to d\gamma$ operator $O_{7\gamma}$.    
Solid circles denote the $b\to d\ell^+\ell^-$ operators $O_{9V}$ and
$O_{10AV}$.}
\end{center}
\end{figure}
\section{The $B\to \ell^+\ell^-\gamma$ amplitude}
\subsection{\label{sec:2}
Direct emission of the real photon from valence quarks of the $B$ meson}
The amplitude of the process when real photon is directly emitted from the 
valence $b$ or $d$ quarks, and the $\ell^+\ell^-$ pair is coupled to the penguin 
is described by diagrams of Fig. \ref{fig:1}. 
This amplitude is equal to 
$\langle \gamma|H_{\rm eff}^{b\to d \ell^+\ell^-}|B\rangle$ with\footnote{
Our notations and conventions are: 
$\gamma^5=i\gamma^0\gamma^1\gamma^2\gamma^3$, 
$\sigma_{\mu\nu}=\frac{i}{2}[\gamma_{\mu},\gamma_{\nu}]$, 
$\varepsilon^{0123}=-1$, $\epsilon_{abcd}\equiv
\epsilon_{\alpha\beta\mu\nu}a^\alpha b^\beta c^\mu d^\nu$, 
$e=\sqrt{4\pi\alpha_{\rm em}}$. } 
\begin{eqnarray}
\label{b2qll}
&&H_{\rm eff}^{b\to d\ell^{+}\ell^{-}}\, =\, 
{\frac{G_{F}}{\sqrt2}}\, {\frac{\alpha_{\rm em}}{2\pi}}\, 
V_{tb}V^*_{tq}\, 
\left[\,-2im_b\, {\frac{C_{7\gamma}(\mu)}{q^2}}\cdot
\bar d\sigma_{\mu\nu}q^{\nu}\left (1+\gamma_5\right )b
\cdot{\bar \ell}\gamma^{\mu}\ell \right.\nonumber\\
&&\left.\qquad\qquad\quad +\, 
C_{9V}^{\rm eff}(\mu, q^2)\cdot\bar d \gamma_{\mu}\left (1\, -\,\gamma_5 \right)   b 
\cdot{\bar \ell}\gamma^{\mu}\ell \, +\, 
C_{10A}(\mu)\cdot\bar d   \gamma_{\mu}\left (1\, -\,\gamma_5 \right) b 
\cdot{\bar \ell}\gamma^{\mu}\gamma_{5}\ell \right]. 
\end{eqnarray} 
The coefficient $C^{\rm eff}_{9V}(\mu, q^2)$ includes long-distance effects related to 
$\bar cc$ resonances in the $q^2$-channel, $q$ the momentum of the $\ell^+\ell^-$ pair \cite{res}.  
The $C_{7\gamma}$ part in Eq. (\ref{b2qll}) emerges from the diagrams
in Fig. \ref{fig:1} with the virtual photon emitted from the penguin 
\begin{equation}
\label{b2qgamma}
H_{\rm eff}^{b\to d\gamma}\, =\,\frac{G_{F}}{\sqrt2}\, V_{tb}V^*_{tq}\, 
C_{7\gamma}(\mu)\,\frac{e}{8\pi^2}\, m_b \cdot
\bar d\,\sigma_{\mu\nu}\left (1+\gamma_5\right )b \cdot F^{\mu\nu}.
\end{equation}
The $B\to \gamma$ transition form factors of the basis operators in (\ref{b2qll})
are defined according to \cite{mk2003}
\begin{eqnarray}
\label{real}
\langle
  \gamma (k,\,\epsilon)|\bar d \gamma_\mu\gamma_5 b|B(p) 
\rangle 
&=& i\, e\,\epsilon^*_{\alpha}\, 
\left ( g_{\mu\alpha} \, pk-p_\alpha k_\mu \right )\,\frac{F_A(q^2)}{M_B}, 
\nonumber
\\
\langle
  \gamma(k,\,\epsilon)|\bar d\gamma_\mu b|B(p)
\rangle
&=& 
e\,\epsilon^*_{\alpha}\,\epsilon_{\mu\alpha\xi\eta} p_{\xi}k_{\eta}\, 
\frac{F_V(q^2)}{M_B},   
\\
\langle
  \gamma(k,\,\epsilon)|\bar d \sigma_{\mu\nu}\gamma_5 b|B(p) 
\rangle\, (p-k)^{\nu}
&=& 
e\,\epsilon^*_{\alpha}\,\left[ g_{\mu\alpha}\,pk- p_{\alpha}k_{\mu}\right ]\, 
F_{TA}(q^2, 0), 
\nonumber
\\
\langle
\gamma(k,\,\epsilon)|\bar d \sigma_{\mu\nu} b|B(p) 
\rangle\, (p-k)^{\nu}
&=& 
i\, e\,\epsilon^*_{\alpha}\epsilon_{\mu\alpha\xi\eta}p_{\xi}k_{\eta}\, 
F_{TV}(q^2, 0).
\nonumber 
\end{eqnarray}
We treat the penguin form factors $F_{TV,TA}(q_1^2,q_2^2)$ as functions of two variables: 
$q_1$ is the momentum of the photon emitted from the penguin, and    
$q_2$ is the momentum of the photon emitted from the valence quark of the 
$B$ meson. Usually one denotes $F_{TV,TA}(q^2,0)\equiv F_{TV,TA}(q^2)$. 
The form factors $F_{A,V,TA,TV}(q^2,0)$ were studied in detail in \cite{mk2003}. 
The parametrizations for these form factors from \cite{mk2003} which satisfy all known 
constraints in the limit $m_b\to\infty$ \cite{korch,bb} will be used in our analysis.  

The amplitude corresponding to diagrams of Fig 1 takes the form \cite{mk2003}:
\begin{eqnarray}
\label{mainmatrix}
A_\mu^{(1)}&=&\langle\gamma (k,\,\epsilon),\,\ell^+(p_1),\,\ell^-(p_2)\left |
H_{\rm eff}^{b\to d\ell^+\ell^-} \right|B(p) \rangle\, =\,
\frac{G_F}{\sqrt{2}}\, V_{tb}V^*_{tq}\,\frac{\alpha_{\rm em}}{2\pi}\, e\, 
\epsilon^*_{\alpha}\nonumber \\
&& \times\left [ 
\frac{2\, C_{7\gamma}(\mu)}{q^2}\, m_b\, 
\left (\varepsilon_{\mu\alpha\xi\eta}p_{\xi}k_{\eta}\,F_{TV}(q^2, 0)\, -\, 
i\left (g_{\mu\alpha}\, pk\, -\, p_{\alpha}k_{\mu}\right )\, F_{TA}(q^2, 0) 
\right)\,\bar\ell (p_2)\gamma_{\mu}\ell (-p_1)
   \right .\nonumber \\
&& \quad C^{\rm eff}_{9V}(\mu, q^2)
\left (\varepsilon_{\mu\alpha\xi\eta}p_{\xi}k_{\eta}\,
                \frac{F_V(q^2)}{M_B}\, -\, 
i\left (g_{\mu\alpha}\, pk\, -\, p_{\alpha}k_{\mu}\right )\,
                \frac{F_A(q^2)}{M_B} 
\right)\,\bar\ell (p_2)\gamma_{\mu}\ell (-p_1)\, + \nonumber \\
&& \quad\left.
C_{10A}(\mu)\left (\varepsilon_{\mu\alpha\xi\eta}p_{\xi}k_{\eta}\,
                \frac{F_V(q^2)}{M_B}\, -\, 
i\left (g_{\mu\alpha}\, pk\, -\, p_{\alpha}k_{\mu}\right )\,
                \frac{F_A(q^2)}{M_B} 
\right)\,\bar\ell (p_2)\gamma_{\mu}\gamma_5\ell (-p_1)
   \right . \Biggr]. 
\end{eqnarray}

\begin{figure}[th]
\begin{center}
\epsfig{file=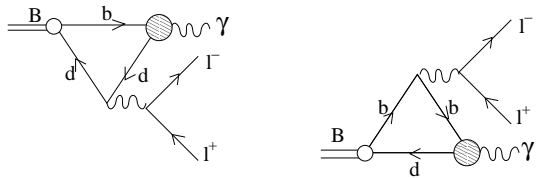,width=7.0cm}
\caption{\label{fig:diagr} Diagrams contributing to 
$B_d\to\ell^+\ell^-\gamma$ discussed in section \protect\ref{sec:3}.
Dashed circles denote the $b\to d\gamma$ operator $O_{7\gamma}$.}
\end{center}
\end{figure}
\subsection{\label{sec:3}Direct emission of the virtual photon from valence quarks of the $B$ meson}
This process is described by diagrams of Fig. 2: the real photon is
emitted from the penguin, whereas one of the valence quarks directly 
emits the virtual photon which then goes into the final $\ell^+\ell^-$ pair. 
The corresponding amplitude  $A^{(2)}_\mu$ has the same structure as the $C_{7\gamma}$ part 
of the amplitude $A^{(1)}_\mu$ with $F_{TA,TV}(q^2,0)$ replaced by $F_{TA,TV}(0,q^2)$. 
The form factors $F_{TA,TV}(0,q^2)$ at the necessary timelike momentum transfers are not known. 
The difficulty with these form factors comes from neutral light vector mesons,  
$\rho^0$ and $\omega$ for $B_d$ decay and $\phi$ for $B_s$ decay, 
which appear in the physical $B\to \gamma \ell^+\ell^- $ decay region. 
These resonances emerge in the amplitude of the subprocess when the photon is emitted 
from the light valence $d$ or $s$ quark. 
We obtain the form factors $F_{TA,TV}(0,q^2)$ for $q^2>0$ 
using gauge-invariant version \cite{otto} of the vector meson
dominance \cite{vmd}. This allows us to express 
the form factors $F_{TA,TV}(0,q^2)$ in terms of the $B\to V$
transition form factors at zero momentum transfer and leptonic constants of vector mesons 
\begin{eqnarray}
\label{zamena1a}
&&F_{TV,TA}(0, q^2)\, =\, F_{TV,TA}(0, 0)\, -\,\sum_V\,2\,f_V g^{B\to V}_+(0)
\frac{q^2/M_V}{q^2\, -\, M^2_V\, +\, iM_V\Gamma_V},
\end{eqnarray}
$M_V$ and $\Gamma_V$ being the mass and width of the vector meson resonance. 
The $B\to V$ transition form factors are defined according to relations 
\begin{eqnarray}
\langle V(q, \varepsilon)|\bar d\sigma_{\mu\nu} b|
B(p)\rangle
\, =\, i\varepsilon^{*\alpha}\,\epsilon_{\mu\nu\beta\gamma}
\left [ 
g^{B\to V}_+(k^2)g_{\alpha\beta}(p+q)^{\gamma} + 
g^{B\to V}_-(k^2)g_{\alpha\beta}k^{\gamma} + 
g^{B\to V}_0(k^2)p_{\alpha}p^{\beta}q^{\gamma}
\right ]. 
\end{eqnarray}
The leptonic decay constant of a vector meson is given by
\begin{eqnarray}
\langle 0|\bar d \gamma_\mu d|V(\varepsilon, p)\rangle=\varepsilon_\mu M_V f_V.  
\end{eqnarray}
We shall use the form factors $g^{B\to V}$ calculated in \cite{ms} using the 
relativistic dispersion approach \cite{m}. 

Notice that the resonance contribution discussed in this Section is similar to the 
resonance contribution in the Wilson coefficient $C^{\rm eff}_{9V}$ which emerge due to 4-quark 
operators. In the case of $C^{\rm eff}_{9V}$ vector mesons containing $\bar cc$ and 
$\bar uu$ pairs 
contribute. 
\begin{figure}[b]
\begin{center}
\epsfig{file=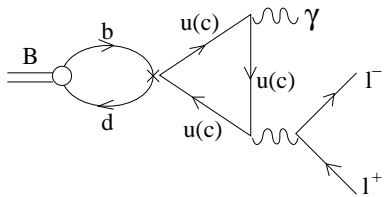,width=5.0cm}
\caption{\label{fig:3} Weak annihilation diagrams discussed in section \protect\ref{sec:4}.
Triangle diagrams with $u$ and $c$ quarks in the loop should be taken into account.}
\end{center}
\end{figure}

\subsection{\label{sec:4}Weak annihilation}
The weak annihilation amplitude $A^{\rm WA}$ is given by triangle diagrams of Fig 3. 
One should take into account $u$ and $c$ quarks in the loop. 
The vertex describing the $\bar bd\to \bar UU$ 
transition ($U=u,c$) reads
\begin{eqnarray}
\label{heffwa}
H_{\rm eff}^{B\to\bar UU} = -\,\frac{G_F}{\sqrt{2}}\, a_1\,V_{Ub}V^*_{Ud}
\,\bar d\gamma_{\mu}(1 -\gamma_5)b 
\,\bar U\gamma_{\mu}(1 -\gamma_5)U, 
\end{eqnarray}
with $a_1\, =\, C_1\, +\, C_2/N_c$, $N_c$ number of colors \cite{stech}. For $N_c\, =\, 3$ 
one finds $a_1\, =\, -0.13$. We now have to take 
\begin{eqnarray}
\langle \ell^+\ell^-\gamma|H_{\rm eff}^{B\to\bar UU}|B\rangle.   
\end{eqnarray}
The $\bar UU$ contribution to this amplitude can be written as 
\begin{eqnarray}
A_\mu^{WA}(\bar UU)=\frac{G_F}{\sqrt{2}}{V_{Ub}V^*_{Ud}}a_1 2e^3
\epsilon_{\mu \varepsilon^* q k}\frac{G_{\gamma\gamma}(M_B^2,k^2=0,q^2|m^2_U)}{q^2}\,
\bar \ell^+ \gamma_\mu \ell^-, 
\end{eqnarray}
where the form factor $G(p^2,k^2,q^2|m^2_U)$ is defined as follows \cite{ms1}
\begin{eqnarray}
\langle \gamma^*(k)\gamma^*(q)|\partial^\nu (\bar U \gamma_\nu \gamma_5 U)|0\rangle
=
e^2\varepsilon^{*\alpha}(k)\varepsilon^{*\beta}(q)
\epsilon_{\alpha\beta kq} G_{\gamma\gamma}(k^2,q^2,p^2|m^2_U).
\end{eqnarray}
For massless $u$-quark in the loop, axial anomaly \cite{abj} fixes
the form factor  
$G_{\gamma\gamma}(p^2,k^2,q^2|0)=-\frac{2N_c (Q_U)^2}{4\pi^2}$. 
For $c$-quark there is additional $q^2$-dependent contribution given by the amplitude  
$m_c \langle \gamma^*(k)\gamma^*(q)|\bar c \gamma_5 c)|0\rangle \sim
m_c^2/M_B^2$, which contains $\psi$ and
$\psi'$ resonances at $q^2>0$. The latter contribution is numerically
negligible compared to contributions discussed in the previous
sections for all $q^2$ in the reaction of interest. 
Therefore, we have 
\begin{eqnarray}
A_\mu^{WA}=-\frac{G_F}{\sqrt{2}}\alpha_{\rm em}e\, a_1
\{V_{ub}V^*_{ud}+V_{cb}V^*_{cd}\}  
\frac{16}{3}
\epsilon_{\mu \varepsilon^* q k}\frac{1}{q^2}\,\bar \ell^+ \gamma_\mu \ell^-.  
\end{eqnarray}
The anomalous contribution is enhanced at small $q^2$, but even here it is suppressed 
by a power of a heavy quark mass compared to the contributions discussed in the previous sections
\cite{bb,wa}. 


\subsection{\label{sec:5}Bremsstrahlung}
Fig. 4 gives diagrams describing Bremsstrahlung. The corresponding contribution to the 
$B\to \ell^+\ell^-\gamma$ amplitude  reads 
\begin{eqnarray}
\label{bremsstrahlung}
A_\mu^{\rm Brems}=i\, e\,\frac{G_F}{\sqrt{2}}\,\frac{\alpha_{\rm em}}{2\pi}\, V^*_{td}V_{tb}\, 
\frac{f_{B_q}}{M_B}\, 2\hat m_{\ell}\, C_{10A}(\mu)\, 
\bar\ell (p_2)
\left [
\frac{(\gamma\epsilon^*)\,(\gamma p)}{\hat t-\hat m^2_{\ell}}\, -\, 
\frac{(\gamma p)\,(\gamma\epsilon^*)}{\hat u-\hat m^2_{\ell}}
\right ]
\gamma_5\,\ell (-p_1). 
\end{eqnarray}

\begin{figure}[h]
\begin{center}
\begin{tabular}{cc}
\epsfig{file=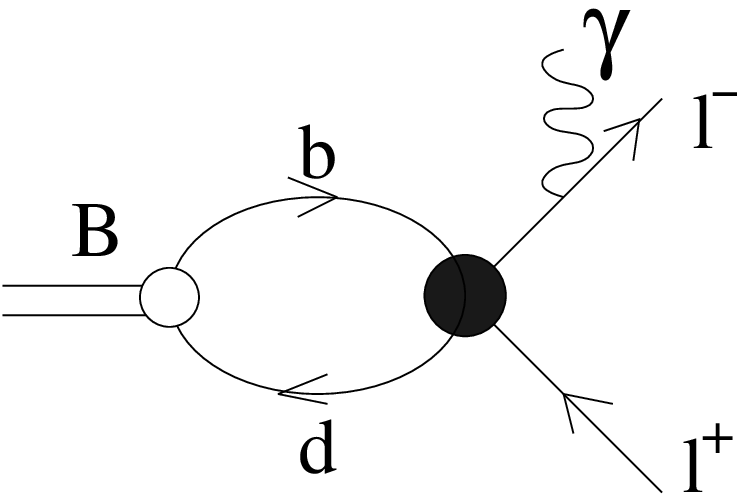,width=3.5cm}
&
\qquad\qquad\epsfig{file=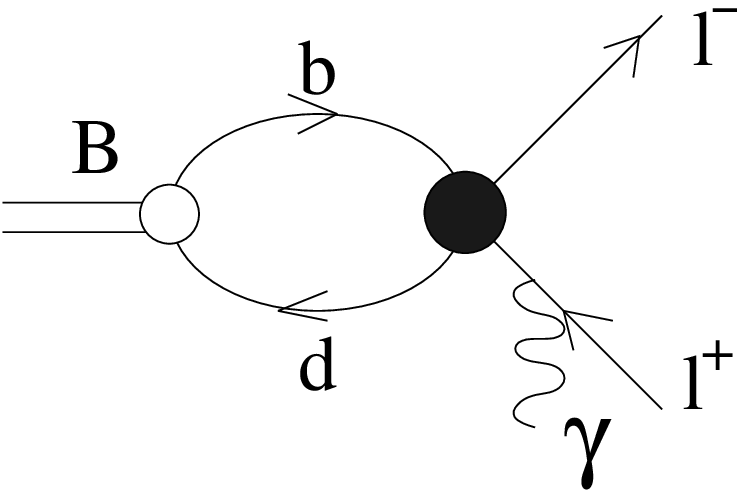,width=3.5cm}
\end{tabular}
\caption{\label{fig:4} Diagrams describing photon Bremsstrahlung. Solid
circles denote the operator $O_{10A}$.}
\end{center}
\end{figure}

\newpage
\section{\label{sec:6}The $B\to \ell^+\ell^-\gamma$ cross-section}
The amplitudes discussed in Sections \ref{sec:2}-\ref{sec:4} have the same Lorentz structure, 
whereas the structure of the Bremsstrahlung amplitude (Section \ref{sec:5}) is
different. 
Therefore, it is convenient to 
write the cross-section as the sum of three contributions: square of the amplitude $A^{1+2+WA}$
which we denote $\Gamma^{(1)}$, square of the amplitude $A^{\rm Brems}$ 
which we denote $\Gamma^{(2)}$, and their mixing, denoted as $\Gamma^{(12)}$:  
\begin{eqnarray}
\label{Gamma1}
&&\frac{d^2\Gamma^{(1)}}{d\hat s\, d\hat t}\, =\, 
\frac{G^2_F\,\alpha^3_{em}\, M^5_1}{2^{10}\,\pi^4}\, 
\left |V_{tb}\, V^*_{tq} \right |^2
\left [ 
x^2\, B_0\left (\hat s,\,\hat t\right )\, +\,
x\,\,\xi\left (\hat s,\hat t\right )\,\tilde B_1\left (\hat s,\,\hat t\right )
\, +\,  
\xi^2\left (\hat s,\hat t\right )\,\tilde B_2\left (\hat s,\,\hat t\right ) 
\right ], 
\\
\label{BiFi}
&&\qquad\qquad B_0\left (\hat s,\,\hat t\right )\, =\,
    \left (\hat s\, +\, 4\hat m^2_{\ell} \right )
    \left (F_1\left(\hat s\right )\, +\, F_2\left(\hat s\right )\right)\, -\, 
    8\hat m^2_{\ell}\,\left |C_{10A}(\mu)\right |^2
    \left (F^2_V\left(q^2\right )\, +\, F^2_A\left(q^2\right )\right), 
    \nonumber \\
&&\qquad\qquad \tilde B_1\left (\hat s,\,\hat t\right )\, =\,
     8\,\left [
                \hat s\, F_V(q^2)\, F_A(q^2)\, 
                Re\left (C^{eff\, *}_{9V}(\mu, q^2)\, C_{10A}(\mu)\right )\, 
        \right .\nonumber\\
&&  \qquad\qquad\qquad\qquad\left . +\,  
                \hat m_b\, F_V(q^2)\, Re\left (C^*_{7\gamma}(\mu)\, 
                \bar F^*_{TA}(q^2)\, C_{10A}(\mu)\right )
              + \hat m_b\, F_A(q^2)\, Re\left (C^*_{7\gamma}(\mu)\, 
                \bar F^*_{TV}(q^2)\, C_{10A}(\mu)\right ) 
        \right ],\nonumber \\
&& \qquad\qquad\tilde B_2\left (\hat s,\,\hat t\right )\, =\,\hat s\, 
    \left (F_1\left(\hat s\right )\, +\, F_2\left(\hat s\right )\right),
   \nonumber\\
&& \qquad\qquad F_1\left (\hat s\right )\, =\, 
   \left (\left |C^{\rm eff}_{9V}(\mu, q^2) \right |^2\, +\,
   \left |C_{10A}(\mu) \right |^2  \right)F^2_V(q^2)
   \, +\,
   \left (\frac{2\hat m_b}{\hat s}\right )^2
   \left |C_{7\gamma}(\mu)\, \bar F_{TV}(q^2)\right |^2\nonumber\\
&& \qquad\qquad \qquad\qquad +\,\frac{4\hat m_b}{\hat s}\, F_V(q^2)\, 
   Re\left (C_{7\gamma}(\mu)\, \bar F_{TV}(q^2)\, C^{eff\, *}_{9V}(\mu, q^2) 
     \right ),\nonumber\\
&& \qquad\qquad F_2\left (\hat s\right )\, =\, 
   \left (\left |C^{\rm eff}_{9V}(q^2, \mu) \right |^2\, +\,
   \left |C_{10A}(\mu)\right |^2  \right)F^2_A(q^2)\, +\,
   \left (\frac{2\hat m_b}{\hat s}\right )^2
   \left |C_{7\gamma}(\mu)\, \bar F_{TA}(q^2)\right |^2\nonumber\\
&& \qquad\qquad\qquad\qquad +\,\frac{4\hat m_b}{\hat s}\, F_A(q^2)\, 
   Re\left (C_{7\gamma}(\mu)\, \bar F_{TA}(q^2)\, C^{eff\, *}_{9V}(\mu, q^2) 
     \right ).\nonumber  
\\
\label{Gamma2}
&&\frac{d^2\Gamma^{(2)}}{d\hat s\, d\hat t} =  
\frac{G^2_F\,\alpha^3_{em}\, M^5_1}{2^{10}\,\pi^4}\, 
\left |V_{tb}\, V^*_{tq} \right |^2\,
\left (\frac{8\, f_{B_q}}{M_B}\right )^2\,\hat m^2_{\ell}\,
\left |C_{10A}(\mu) \right |^2 
\left [
   \frac{\hat s\, +\, x^2/2}
         {(\hat u\, -\,\hat m^2_{\ell})(\hat t\, -\,\hat m^2_{\ell})}\, 
-\,\left (\frac{x\,\hat m_{\ell}}
        {(\hat u\, -\,\hat m^2_{\ell})\, (\hat t\, -\,\hat m^2_{\ell})}
   \right )^2\, 
   \right ]
\\
\label{Gamma12}
&&\frac{d^2\Gamma^{(12)}}{d\hat s\, d\hat t}= 
\frac{G^2_F\,\alpha^3_{em}\, M^5_1}{2^{10}\,\pi^4}\, 
\left |V_{tb}\, V^*_{tq} \right |^2\,\frac{16\, f_{B_q}}{M_B}\, \hat m^2_{\ell}
\,\frac{x^2}{(\hat u\, -\,\hat m^2_{\ell})(\hat t\, -\,\hat m^2_{\ell})}
 \\ 
&&\qquad\times
\left [
\frac{2\, x\, \hat m_b}{\hat s}\, Re\left (C^*_{10A}(\mu)C_{7\gamma}(\mu)
\bar F_{TV}(q^2, 0)\right )\,
 +\, x\, F_V(q^2)\, Re\left (C^*_{10A}(\mu)C^{\rm eff}_{9V}(\mu, q^2)\right )
  +\,\xi(\hat s,\hat t)\, F_A(q^2)\,\left |C_{10A}(\mu) \right |^2 
  \right ]. \nonumber 
\end{eqnarray}
Here
\begin{eqnarray}
\label{mandelstam}
\hat s\, =\,\frac{\left (p\, -\, k\right )^2}{M_B^2},\qquad 
\hat t\, =\,\frac{\left (p\, -\, p_1\right )^2}{M_B^2},\qquad 
\hat u\, =\,\frac{\left (p\, -\, p_2\right )^2}{M_B^2}. 
\end{eqnarray} 
with 
$\hat s\, +\,\hat t\, +\,\hat u\, =\, 1\, +\, 2\hat m^2_{\ell}$, 
$\hat m^2_{\ell}\, =\,  m^2_{\ell}/M^2_B$, 
$\hat m_b\, =\, m_b/M_B$ and 
\cite{mk2003} 
\begin{eqnarray}
x\, =\, 1\, -\, \hat s,\qquad 
\cos\theta\, =\,\frac{\xi\left (\hat s,\hat t\right )}
          {x\,\sqrt{1\, -\, 4\hat m^2_l/\hat s}},\qquad 
\xi\left (\hat s,\hat t\right )\, =\,\hat u\, -\,\hat t.
\end{eqnarray} 
In the above formulas the complex form factors $\bar F_{TV,TA}$ and defines as follows 
\begin{eqnarray}
\bar F_{TV}(q^2)&=&F_{TV}(q^2,0)+F_{TV}(0,q^2)
-\frac{16}{3}\frac{V_{ub}V^*_{ud}+V_{cb}V^*_{cd}}{V_{tb}V^*_{td}}
\frac{a_1}{C_{7\gamma}}\frac{f_B}{m_b},
\nonumber\\
\bar F_{TA}(q^2)&=&F_{TA}(q^2,0)+F_{TA}(0,q^2).
\end{eqnarray} 
The expressions (\ref{Gamma2}) and (\ref{Gamma12}) contain the
infrared pole which requires a cut in the energy of the emitted
photon. Clearly, the contribution of the pole is proportional to the lepton mass.  
\newpage
\section{\label{sec:7}Numerical analysis}
\subsection{Parameters}
For numerical estimates we use the following values: 
\begin{itemize}
\item
The Wilson coefficients at $\mu=5$ GeV corresponding to $C_2(M_W)=\, -1$: \\
$C_1(5\, GeV)=\,0.241$, $C_2(5\, GeV)=\, -1.1$, $C_{7\gamma}(5\, GeV)=\, 0.312$, 
$C_{9V}(5\, GeV)=\, -4.21$, $C_{10A}(5\, GeV)=4.64$ \cite{wils}.
\item
The CKM matrix elements 
$\left |V^*_{tb}V_{td} \right | =(8.3\pm 1.6)10^{-3}$, 
$\left |V^*_{tb}V_{ts} \right | =(4.7\pm 0.8)10^{-2}$ \cite{pdg}. 
\item
The $B$-meson lifetimes 
$\tau(B_d)=1.536\pm 0.014$ ps and 
$\tau(B_s)=1.461\pm 0.057$ ps \cite{pdg}.
\item
For the $B_d\to\gamma$ form factors we use parametrizations from 
\cite{mk2003}
\begin{eqnarray}
\label{frank}
F_{i}(q^2)&=&\beta_i\frac{f_B M_B}{\Delta_i+E^\gamma},\qquad 
E^\gamma=\frac{M_B}{2}\left(  1-\frac{q^2}{M_B^2} \right),
\end{eqnarray} 
with $i=V,A,TV,TA$ and the parameters $\beta$ and $\Delta$ 
given in Table \ref{table:frank}.
\begin{table}[h]
\caption{\label{table:frank}
Parameters for the form factors in Eq. (\ref{frank}) from \cite{mk2003}}
\centering
\begin{tabular}{|l|c|c|c|c|}
\hline
 Parameter              &  $F_V$   & $F_{TV}$   & $F_{A}$   & $F_{TA}$   \\
\hline
$\beta$  (GeV$^{-1}$)  &  0.28    &   0.30   &   0.26     &   0.33    \\   
$\Delta$ (GeV)         &  0.04    &   0.04   &   0.30     &   0.30    \\
\hline
\end{tabular}
\end{table}

\item

The form factors $g^{B\to V}_+$ at $q^2=0$ for $B_d\to\rho^0$,
$B_d\to\omega$ transitions and 
$B_s\to\phi$ are given in Table \ref{table:ff}. The 
$B_s\to\gamma$ and the $B_d\to\gamma$ form factors are taken to
be equal to each other. 

\begin{table}[h]
\caption{\label{table:ff}
The leptonic decay constants $f_V$ and the $B\to V$ transition form factors 
$g^{B\to V}_+$ at $q^2=0$.}
\centering
\begin{tabular}{|c|c|c|c|}
\hline
                 &  $\rho^0$    & $\omega$ & $\phi$ \\
\hline
$f_V$ (MeV)       &  154         &   45.3     & $-58.8$    \\   
\hline
$g^{B^0\to V}_+(0)$  & $0.19$   & $-\, 0.19$  &  $-\, 0.38$ \\
\hline
\end{tabular}
\end{table}
\end{itemize}

\subsection{Results}
\begin{figure}[tbh]
\begin{center}
\begin{tabular}{ccc}
\mbox{\epsfig{file=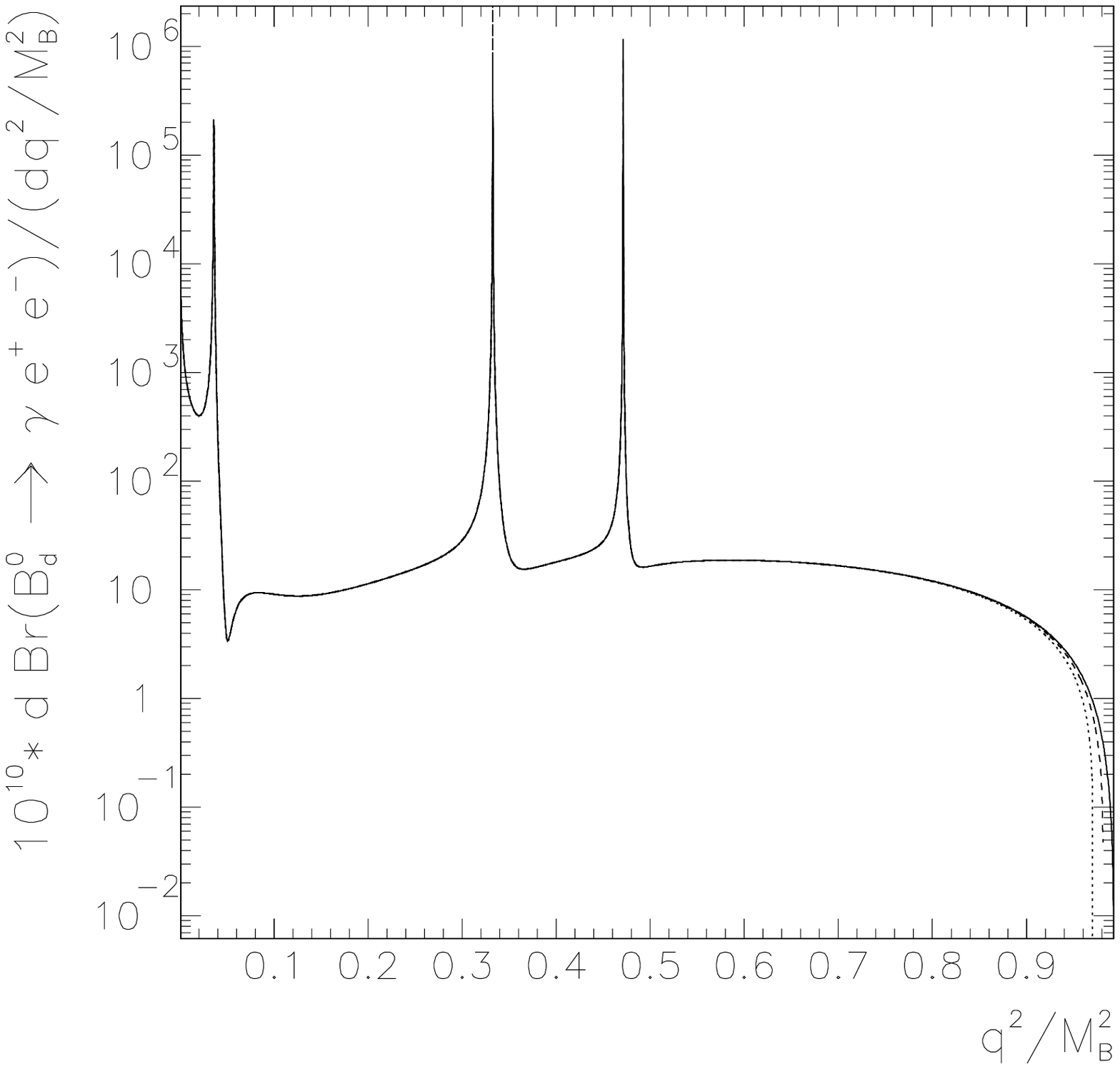,width=5.cm}} &
\mbox{\epsfig{file=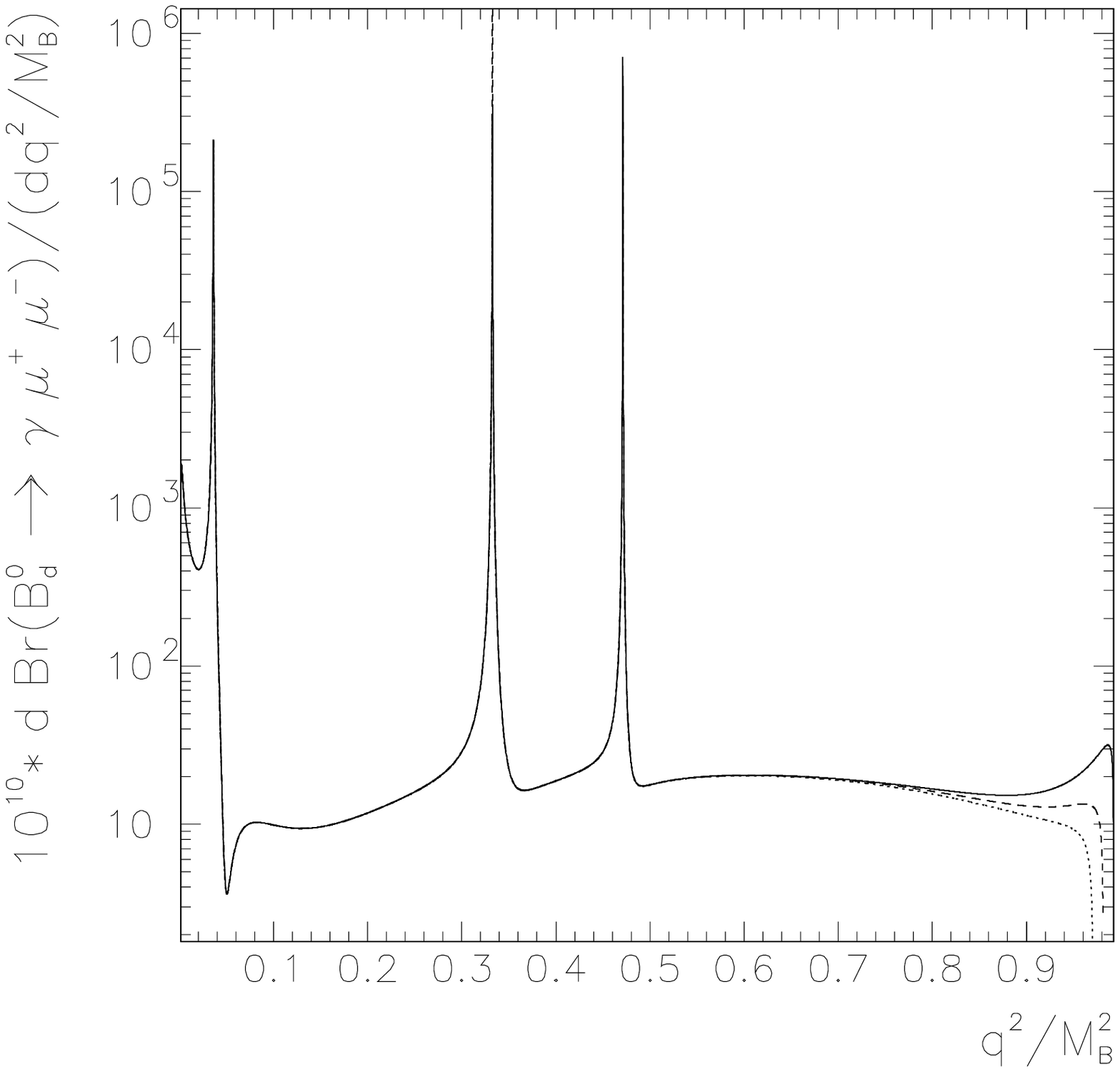,width=5.cm}} &
\mbox{\epsfig{file=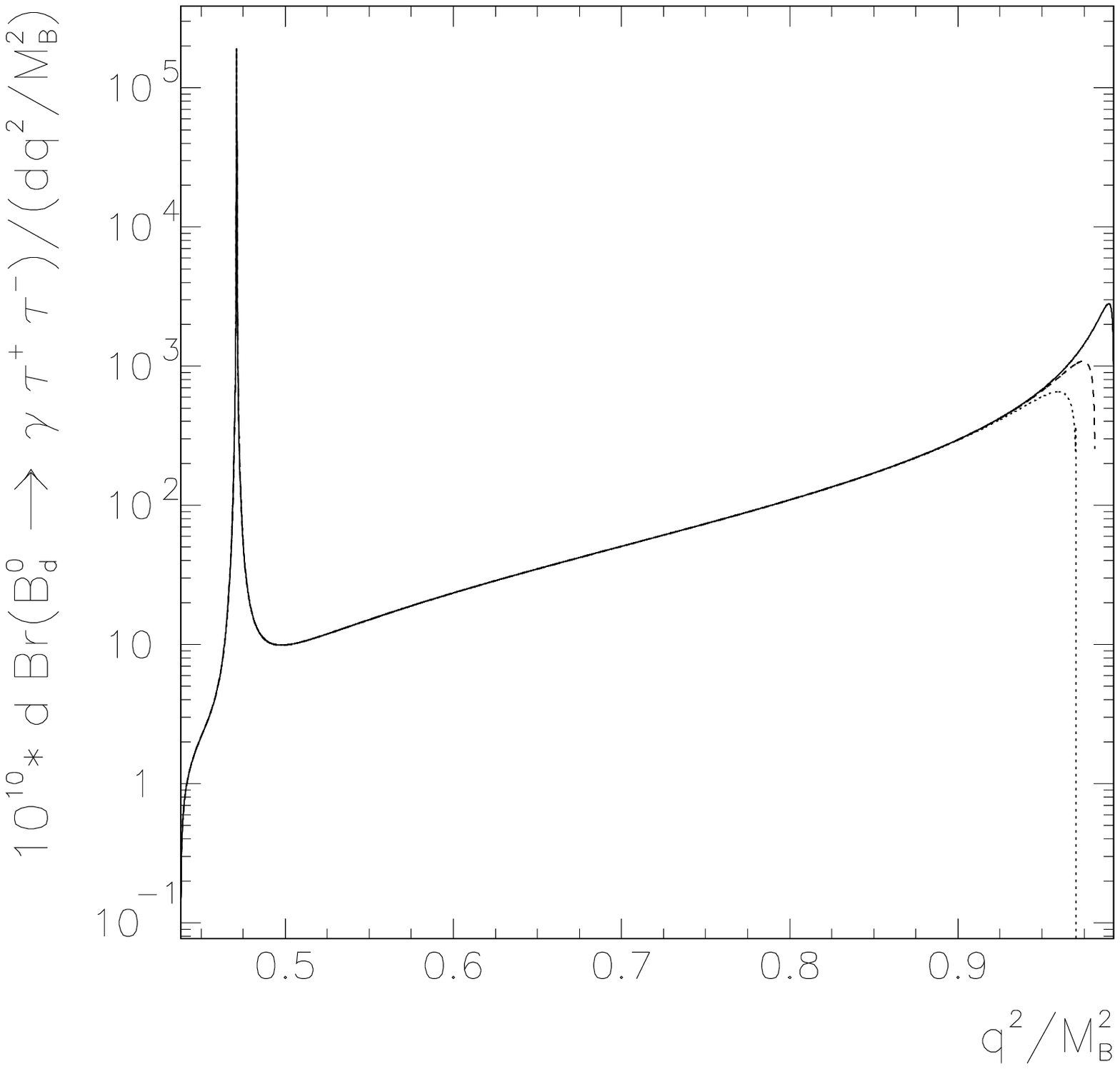,width=5.cm}}
\end{tabular}
\caption{\label{fig:52} 
Dilepton $q^2$-spectrum 
in $B_s\to e^+e^-\gamma$ (left), $B_s\to \mu^+\mu^-\gamma$ (central), 
and $B_s\to \tau^+\tau^-\gamma$ (right) 
for different photon energy cuts: 
$E^{\gamma}_{min}= 20$ MeV (solid),
$E^{\gamma}_{min}= 50$ MeV (dashed),
$E^{\gamma}_{min}= 80$ MeV (dotted).}
\end{center}
\end{figure}

The calculated dilepton differential distributions are shown in
Fig. \ref{fig:52} for different leptons in the final state and 
for different values of the photon energy cut - the minimal photon 
energy in the $B$-meson rest frame $E^{\gamma}_{min}$. 
Table \ref{table:res} gives  the dependence of the  integrated 
$B_{d,s}\to\ell^+\ell^-\gamma$ 
decay rate on  $E^{\gamma}_{min}$. A particular choice of  
$E^{\gamma}_{min}$ depends on the energy resolution in a specific
experiment: namely, 
$E^{\gamma}_{min}=80$ MeV and $E^{\gamma}_{min}=20$ MeV 
in Table \ref{table:res} correspond to the expected accuracy of the 
$B$-meson reconstruction of ATLAS and LHCb, respectively.  
\begin{table}[b]
\caption{\label{table:res}
The $B_{d,s}\to\ell^+\ell^-\gamma$ decay rates as functions of the
minimal photon energy $E^{\gamma}_{min}$. The region of the 
$J/\psi$ and $\psi'$ resonances $0.33\le \hat s \le 0.55$ was excluded 
from the $\hat s$ integration, that corresponds to the experimental procedure 
adopted at LHC \cite{lhc,nrs}.}\centering\begin{tabular}{|c|c|c|c|c|c|c|c|c|c|}
\hline
\protect\( m_{\ell}\protect\) & \multicolumn{3}{|c|} {\protect\( m_e     \protect\)} &
                                \multicolumn{3}{|c|} {\protect\( m_{\mu} \protect\)} &
                                \multicolumn{3}{|c|} {\protect\( m_{\tau}\protect\)} \\
\hline
\protect\( E^{\gamma}_{min}\protect\) (MeV) & 20 & 50 & 80 &
                         20 & 50 & 80 &
                         20 & 50 & 80 \\
\hline
\protect\( Br\left ( B_d\to\ell^+\ell^-\gamma\right )\,\times\, 10^{10}\protect\) [This work] &
                         3.95 & 3.95 & 3.95 &
                         1.34 & 1.32 & 1.31 &
                         3,39 & 2.37 & 1.87 \\
\protect\( Br\left ( B_s\to\ell^+\ell^-\gamma\right )\,\times\, 10^9\protect\) [This work] &
                         24.6 & 24.6 & 24.6 &
                         18,9 & 18.8 & 18.8 &
                         11.6 & 8.10 & 6.42 \\
\hline
\protect\( Br\left ( B_d\to\ell^+\ell^-\gamma\right )\,\times\, 10^{10}\protect\) \protect\cite{kitay} &
                         1.01 & 1.01 & 1.01 &
                         0.66 & 0.62 & 0.61 &
                         3.39 & 2.38 & 1.88 \\
\protect\( Br\left ( B_s\to\ell^+\ell^-\gamma\right )\,\times\, 10^9\protect\) \protect\cite{kitay} &
                         3.30 & 3.29 & 3.29 &
                         2.16 & 2.06 & 2.00 &
                         11.6 & 8.15 & 6.47 \\
\hline
\protect\( Br\left ( B_s\to\ell^+\ell^-\gamma\right )\,\times\, 10^9\protect\) \protect\cite{sehgal} &
                         $20$  & $20$  & 20 &
                         $12$  & $12$  & 12 &
                         $-$  & $-$  & $-$ \\
\hline
\end{tabular}
\end{table}

\begin{figure}[tbh]
\begin{center}
\begin{tabular}{ccc}
\mbox{\epsfig{file=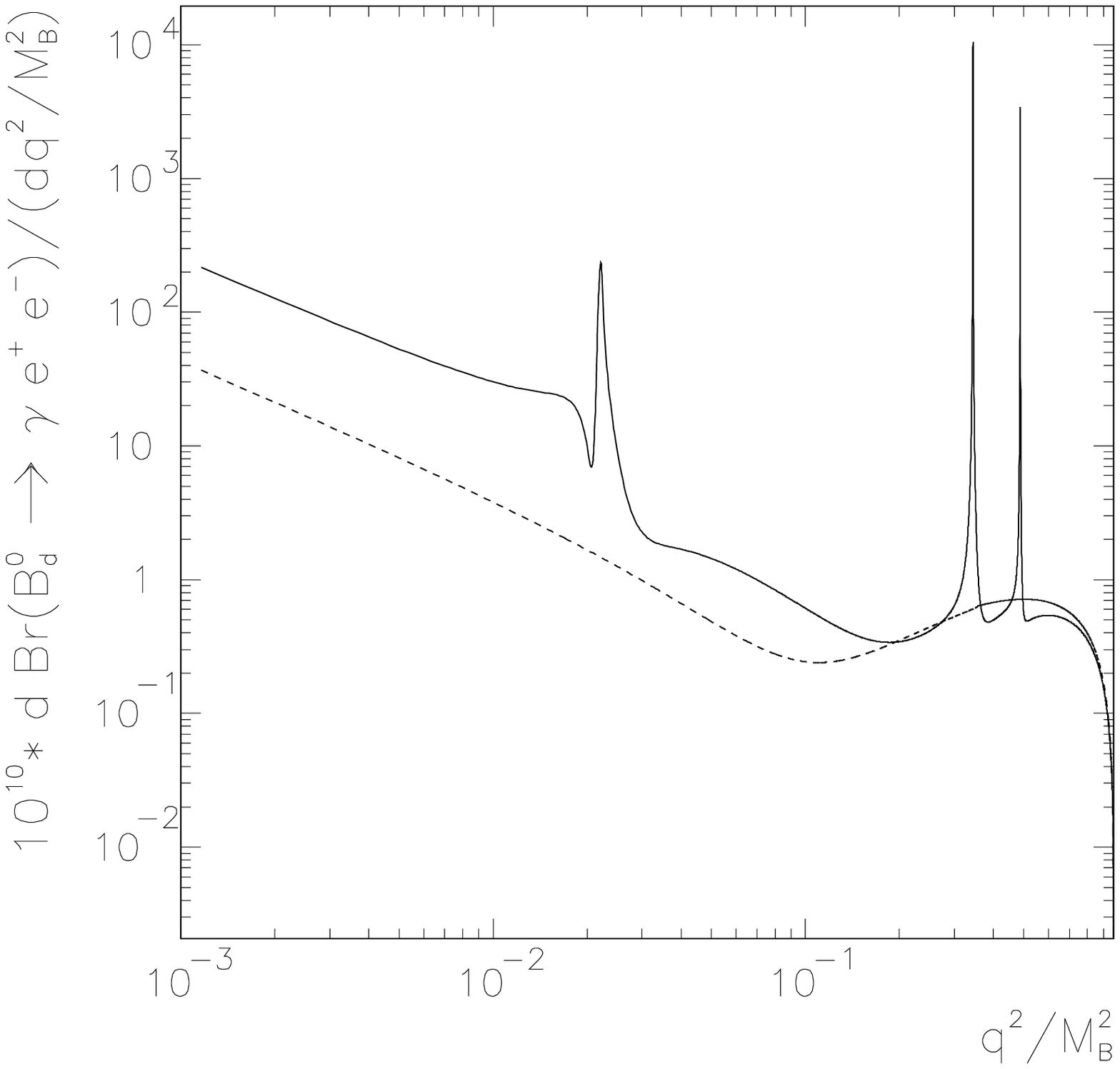,width=5.cm}} &
\mbox{\epsfig{file=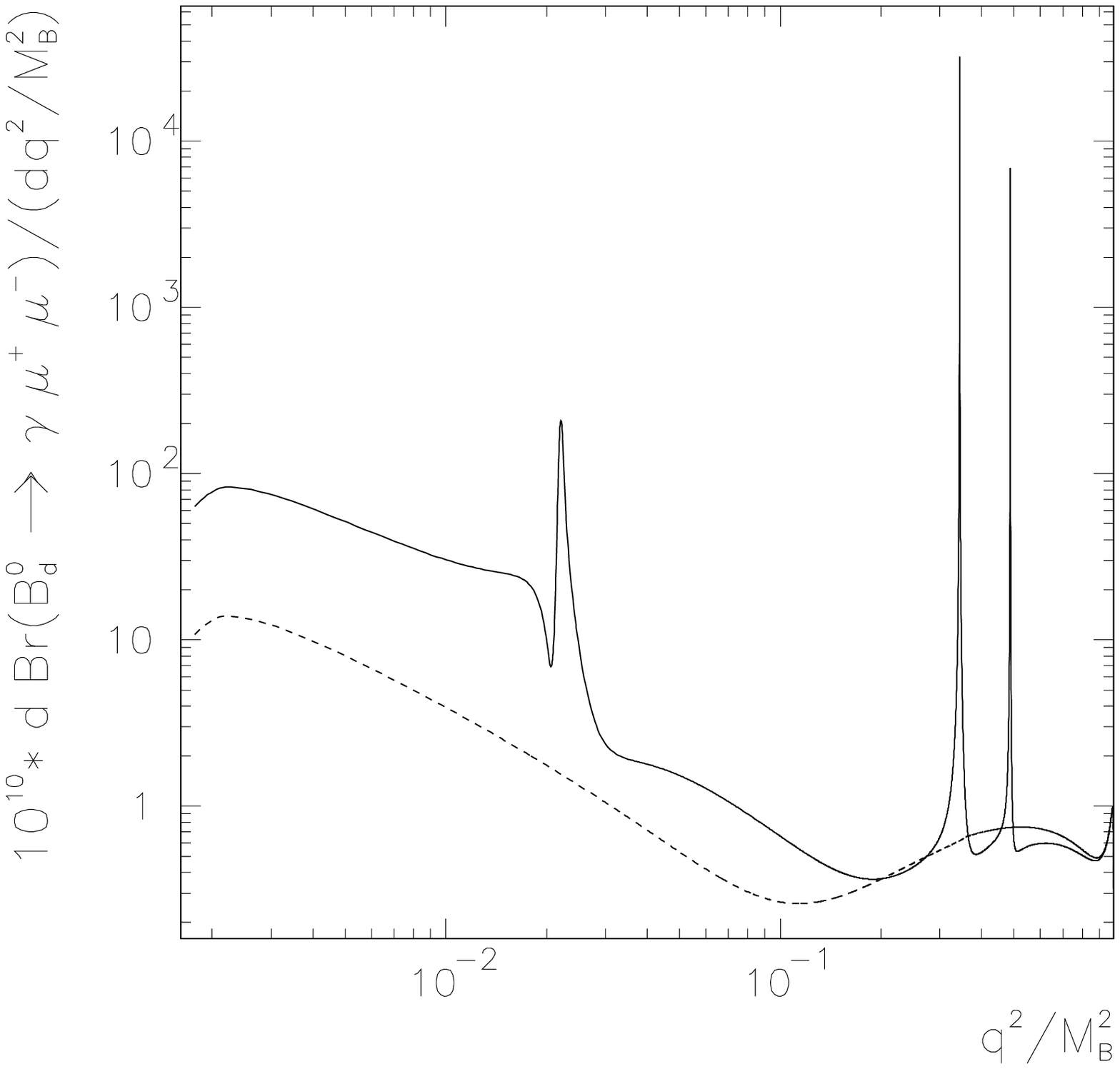,width=5.cm}} &
\mbox{\epsfig{file=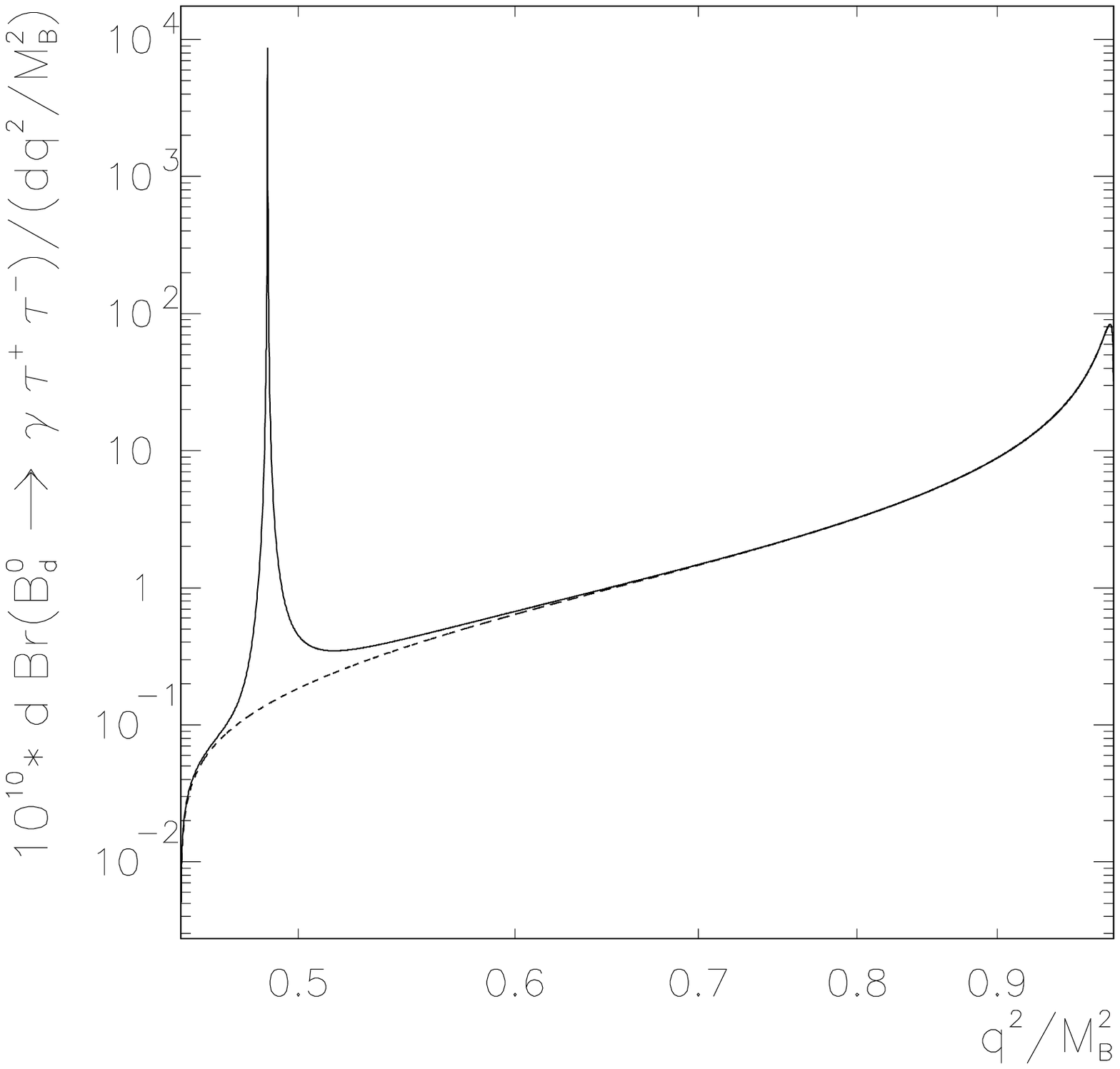,width=5.cm}}
\end{tabular}
\caption{\label{fig:53a} 
Dilepton $q^2$-spectrum in $B_d$ decays: 
$B_d\to e^+e^-\gamma$ (left), $B_d\to \mu^+\mu^-\gamma$ (central), 
and $B_d\to \tau^+\tau^-\gamma$ (right) 
for 
$E^{\gamma}_{min}= 20$ MeV. Solid line - our result; 
dashed line -
result corresponding to \cite{kitay}, where the contribution of the
direct virtual photon emission
from valence quarks of the $B$ meson (Section \ref{sec:3}) was not taken into account.}
\end{center}
\end{figure}

\begin{figure}[tbh]
\begin{center}
\begin{tabular}{ccc}
\mbox{\epsfig{file=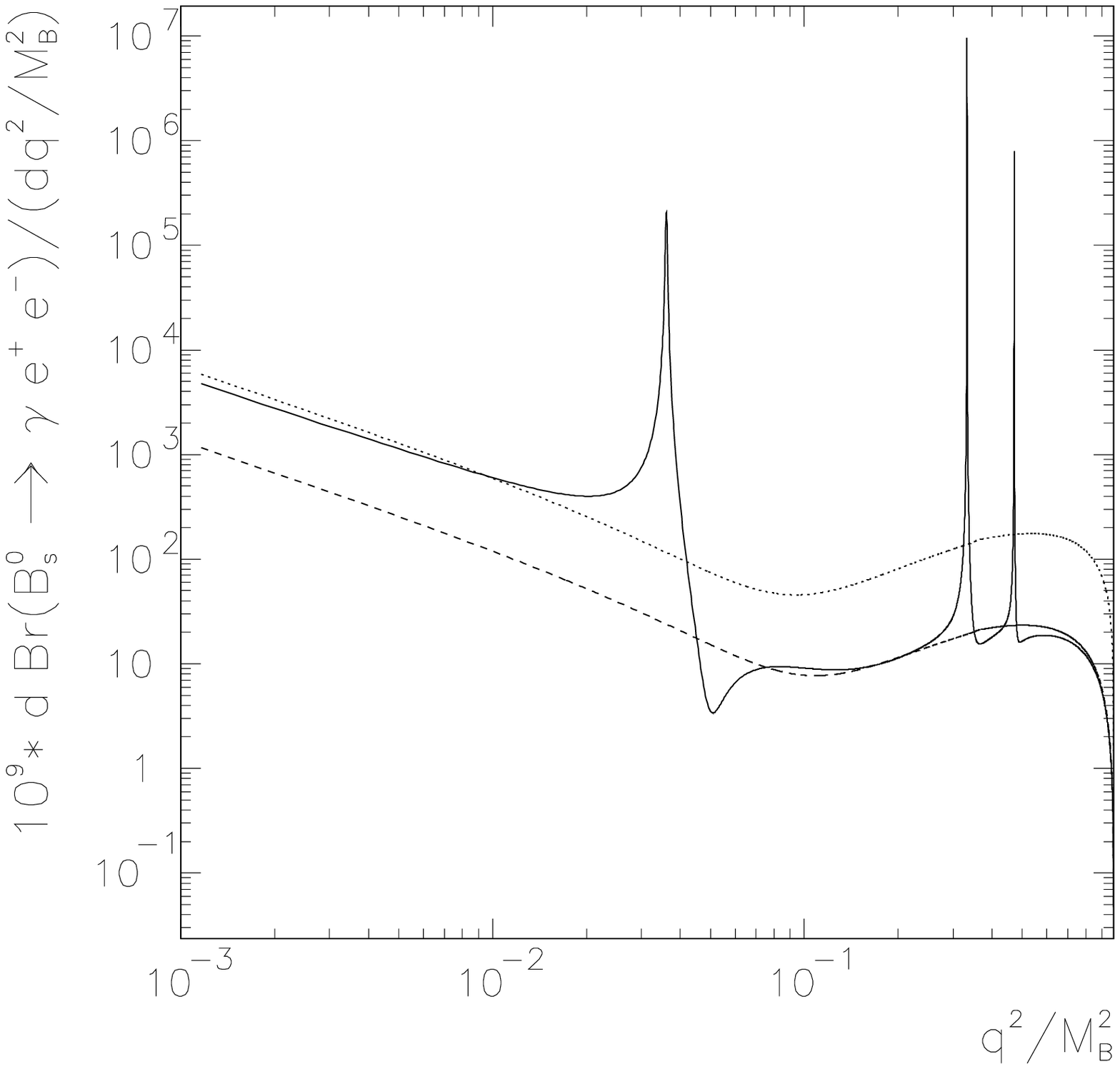,width=5.cm}} &
\mbox{\epsfig{file=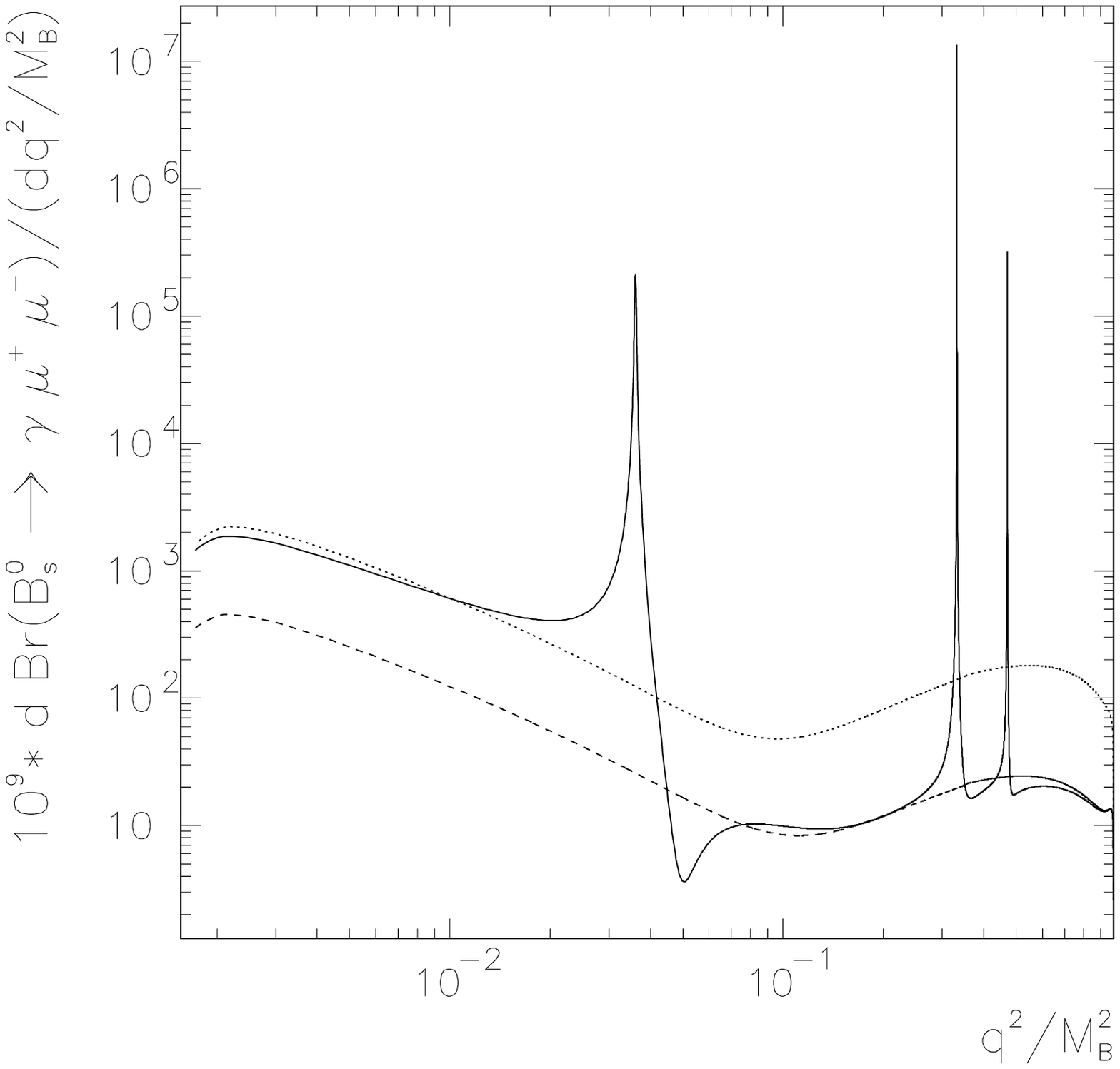,width=5.cm}} &
\mbox{\epsfig{file=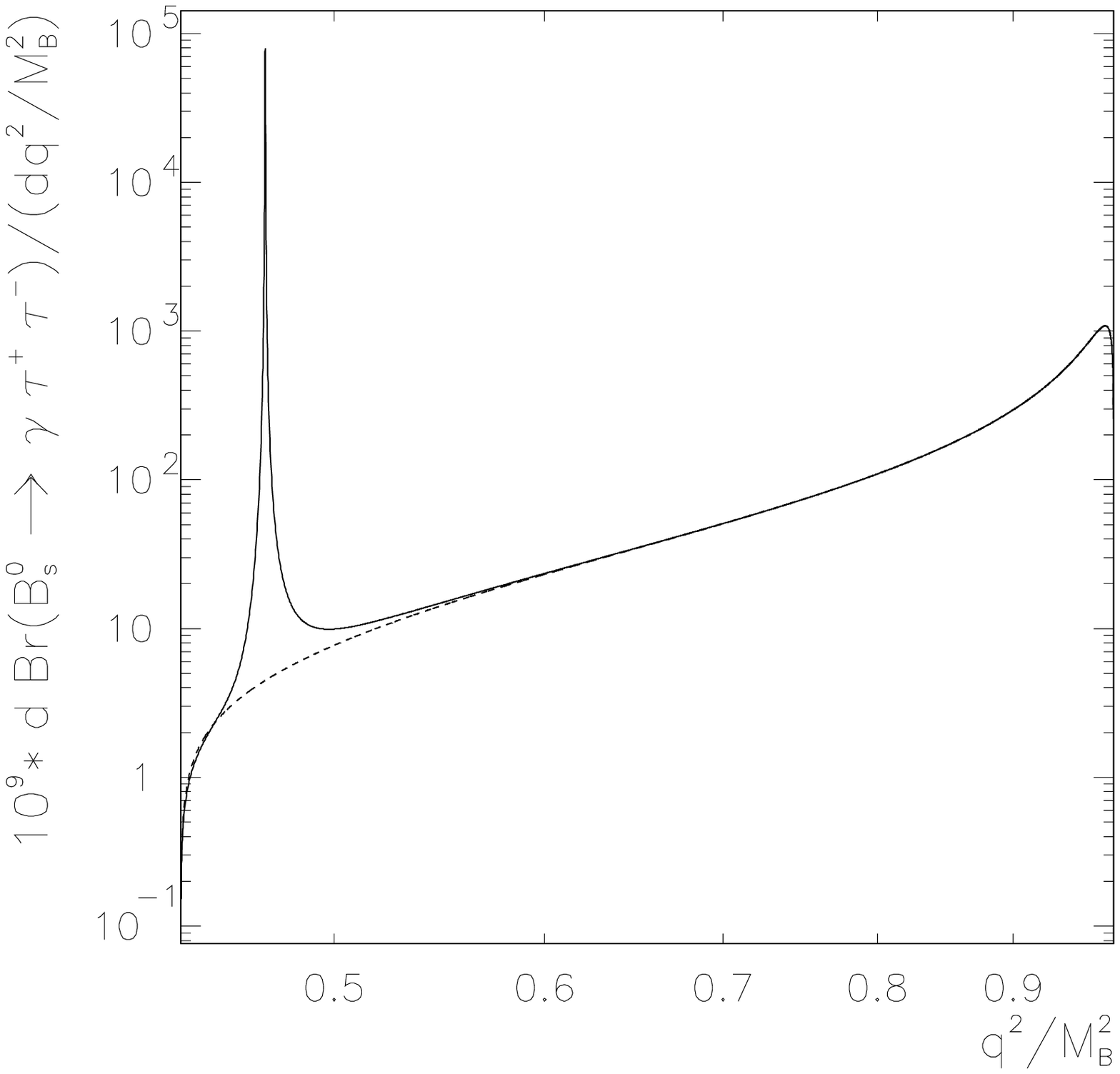,width=5.cm}}
\end{tabular}
\caption{\label{fig:53b} 
Dilepton $q^2$-spectrum in $B_s$ decays: 
$B_s\to e^+e^-\gamma$ (left), $B_s\to \mu^+\mu^-\gamma$ (central), 
and $B_s\to \tau^+\tau^-\gamma$ (right) 
for 
$E^{\gamma}_{min}= 20$ MeV. 
Solid line - our result, 
dashed line -  \cite{kitay}, 
dotted line -  \cite{sehgal}. 
}
\end{center}
\end{figure}

Our results for the {\it integrated} decay rate in Table \ref{table:res} are close to the 
results reported in \cite{sehgal}. 
Notice however that in  \cite{sehgal} resonance contributions were 
not considered and the 
$B\to \gamma$ form factors corresponding to the limit $m_b\to\infty$ were used.  
As shown in \cite{mk2003}, corrections to the asymptotic formulas in $B\to \gamma$ 
form factors at large $q^2$ are rather large, therefore the {\it differential} distributions 
calculated here differ strongly from those of \cite{sehgal}.\footnote{Let us point out that in 
distinction to the $J/\psi$ and $\psi'$, 
the region of light vector resonances will not be excluded from the experimental analysis. 
That is because for small values of $\hat s$ photons have sufficient energies to be
registered in the electromagnetic calorimeter. Moreover, the region of
small $\hat s$ gives the main contribution to the $B\to \mu^+\mu^-\gamma$ signal at LHC.
For $B_s\to e^+e^-\gamma$ and $B_s\to \mu^+\mu^-\gamma$ decays, the $\phi$-meson resonance 
contribution leads to a strong enhancement of the full spectrum compared to the non-resonance
one.}
Our results both for the integrated and the
differential distributions in the $B\to e^+e^-\gamma$ and $B\to
\mu^+\mu^-\gamma$ decays in the region $q^2\lesssim 2$ GeV differ strongly from the 
results of \cite{kitay} where the 
contribution of the direct virtual photon emission from valence quarks of the 
$B$-meson (Section \ref{sec:3}) was not taken into account (Figs. \ref{fig:53a}, \ref{fig:53b}). 

\section{Conclusions}
\label{concl}
We studied the $B_{d, s}\to\ell^+\ell^-\gamma$ decays taking into account photon 
emission from the $b$-quark loop, weak annihilation, and Bremsstrahlung from leptons 
in the final state. 
Special emphasis was laid on long-distance QCD effects in 
$B\to\ell^+\ell^-\gamma$ decays related to the photon 
emission from the $b$-quark loop. The contribution of light 
vector-meson resonances related to the direct virtual photon emission from valence quarks 
of the $B$-meson, not taken into account in previous analyses, was shown 
to be essential for a proper description of the process. In particular, 
this contribution affects stongly dilepton mass spectra in processes 
with light leptons in the final state. 
Reliable predictions for dilepton mass spectrum in $B_{d, s}\to \ell^+\ell^-\gamma$ decays 
within the Standard Model were given. 
\acknowledgments
We would like to thank L.Sehgal for 
clarifications concerning his work and L.Smirnova for stimulating discussions. 
We gratefully acknowledge financial support under a R\"uckkehr Stipendium of 
the Alexander von Humboldt-Stiftung (D.M.) and 
the INTAS grant YSF2001/2-122 (N.N.).

\end{document}